# Self-Assembly of Facetted Particles Triggered by a Moving Ice Front

*Florian Bouville,* [†,‡] *Eric Maire,* [‡] *Sylvain Deville\*,* [†]

[†] Laboratoire de Synthèse et Fonctionnalisation des Céramiques, UMR3080 CNRS/Saint-Gobain, 84306 Cavaillon, France

[‡] Université de Lyon, MATEIS, INSA de Lyon, 69621 Villeurbanne, France

The possibility to align and organize facetted particles in the bulk offers intriguing possibilities for the design and discovery of materials and architectures exhibiting novel functional properties. The growth of ice crystals can be used to trigger the self-assembly of large, anisotropic particles and consequently to obtain three-dimensional porous materials of large dimensions in a limited amount of time. These mechanisms have not been explored so far, due to the difficulty to experimentally investigate these systems. Here we elucidate the self-assembly mechanisms of facetted particles driven by ice growth by a combination of X-ray holo-tomography and discrete element modeling, providing insights into both the dynamics of self-assembly and their final packing. The encapsulation of particles is the result of a delicate balance between the force exerted by the percolating network of concentrated particles and the force exerted by the moving interface. We illustrate the benefits of such self-assembly for thermal management composite materials.

1. Introduction

The controlled alignment of facetted or anisotropic particles holds the promise of a new class of crystalline materials at all length scales, exhibiting appealing structural and functional properties[1]. Understanding and controlling how such particles organize is required for the optimization of functional properties. Superlattices, for instance, with translational and rotational order of the building blocks, are currently investigated for their optical and electrical properties or for plasmonic applications[2].

Most of the approaches developed so far are based on the evaporation of a solvent[3,4] or the sedimentation of particles[5]; ordering in these phenomena is ultimately controlled by the pressure applied to a dense particles assembly, the particle interactions[6] and the rate at which the particle concentration increases. Reaching an optimal packing thus requires a fine control over the particle shape, polydispersity, self-assembly driving force, and particle interactions, which restrict the applicability of the approaches to specific systems.

For larger particles (typically above one micron in diameter for ceramics), the diffusion is limited because of their weight and therefore so is their capacity to form dense structures upon concentration with conventional methods. Sequential deposition methods[7–9] are able to order anisotropic particles, although such methods are limited to thin samples and work only with high aspect ratio particles because the alignment results from shear forces acting on the particles.

Anisotropic structural or functional properties could thus emerge from an assembly of anisotropic building blocks with an approximate alignment. Yet, very few approaches are able to provide such alignment in bulk materials (centimeter scale) and within reasonable time scale (minutes). The growth of ice crystals in a colloidal suspension can be used to drive the self-assembly of anisotropic building blocks[10–12]. Several benefits are associated with ice growth in this case, such as the possibility to obtain bulk materials or a reduced processing time of the order or minutes –vs. hours or days for evaporation or sedimentation processes[5]. In addition, the process is in principle almost independent of the nature of the materials used[13].

Although alignment has been demonstrated experimentally, the underlying phenomena have not been deeply investigated so far. Several mechanisms have been hypothesized[10,12], and numerous important questions must be addressed. What is the driving force for the self-assembly? How do particles organize? How should the freezing conditions be adjusted to optimize the structural organization of the particles? How much do particle-particle interactions matter? The answers to these questions are essential for a proper control of the phenomenon and the resulting functional properties of the materials.

We demonstrate here how the ice growth induced self-assembly of facetted particles can be accurately described and understood by a combination of discrete element modeling and observation of the experimental spatial distribution using high resolution X-ray computed holo-tomography. These tools are used to explore the self-assembly behavior and predict the evolution of such colloidal system. The benefits in terms of functional properties of the resulting structures are illustrated with the thermal properties of Boron Nitride (BN)/silicon rubber composites. The methodology developed here is then extended to predict the self-assembly of cubic particles.

2. Materials and Methods

2.1 Materials processing

Suspensions were prepared by mixing distilled water, a cellulose ether (Tylose H4000P2, Shin Etsu, at a weight ratio Tylose on BN powder of 0.5%) an organic binder (Poly Ethylene Glycol 4M, Sigma Aldrich, 2 wt.% of dried powder mass) and the BN powder (TRES BN PUHP 3008,

Saint-Gobain BN products) at different volume fractions. The powder used in the study present platelet morphology with a diameter around 8 $\mu$m and 1 $\mu$m in thickness. After a first mixing step with a propeller stirrer, ultra sonic mixing was conducted with a sonotrode (Digital Sonifier 250) with an applied energy around 150 W.h/kg.

Freezing of the slurries was done by pouring them into a silicone mold (diameter 20 mm, 21 mm height) placed on a cooled copper plate. The copper plate was cooled by silicone oil at a temperature regulated by a cryothermostat. The cooling rate were adjusted between 0.5°C/min and 2°C/min. Faster cooling rate were obtain by dipping a copper finger with the mold on top in liquid nitrogen. Measurement of ice front velocity was made by dipping a ruler at different times in the slurry and measuring the height of frozen sample. Once freezing was completed, the samples are freeze-dried for at least 48 hrs in a commercial freeze-dryer (Free Zone 2.5 Plus, Labconco, Kansas City, Missouri, USA), to ensure a complete removal of the ice crystals.

Holo-tomography experiments were carried out at European Synchrotron Facility at the beam line ID-22. More details on this technique can be found elsewhere[14]. To enhance phase contrast, porous sample were infiltrated by silicone rubber under vacuum prior to observation. Segmentation and density measurements were made using the software Fiji[15].

2.2 Thermal diffusivity measurement

A flash lamp method was used to measure the thermal diffusivity of the BN/Silicone rubber composite. Disk samples of 20 mm diameter and 5 mm thick were put in a holder, a flash produced by a lamp heated a side of the sample and the temperature profile was recorded on the other side. The thermal diffusivity was obtained by a fit of this profile by the quadripole method[16]. At least five different measurements were made on each sample.

2.3 Percolation threshold estimation

The percolation threshold of platelet particles can be estimated from an empirical formula developed for permeability clay composite[17,18]:

$$\phi_p = \frac{2}{3S+1}\frac{W}{L}p_c$$

With $W$ and $L$ the thickness and length of the platelet respectively, $p_c = 0.718$ and $S$ the order parameter of platelet orientation given by the following relationship:

$$S = \frac{3<cos^2\theta> - 1}{2}$$

Where $<cos^2\theta>$ is the mean of the platelet angles with respect to the preferred direction (equal to 54° for random orientation).

2.4 Discrete Elements Modeling

Discrete element modeling was carried out by a custom version of the LAMMPS Package. To reproduce the lamellar structure characteristic of ice crystal growth, a rectangular simulation box was created. The method used to produce platelet-like particles using spherical ones consist in the creation of a cylindrical region in the simulation box. The region was then filled with spheres with a diameter corresponding to the thickness of the particle, here 1 $\mu$m. During the simulation, the spheres were fixed together as a rigid body, the software just sums the interaction on all the spheres to calculated the resulting displacement and torque applied to the assembly. To introduce the polydispersity of particle size, a random number from a Gaussian statistical distribution was used to adjust the radius of each cylindrical region. The distribution was centered on 4 $\mu$m with standard deviation of 1 $\mu$m. To avoid any overlap of the region that could lead to truncated particles, they were created successively at different x coordinate, while the z and y coordinates was randomly selected. The simulation box dimensions were first set to 300 x 200 x 200 $\mu m^3$, thus a dilute system was obtain. The box sizes were then shrunk to dimensions of 100 x 50 x 50 $\mu m^3$. The number of particles was calculated to obtain the desired volume fraction, knowing the density of BN (2.26 g/cm$^3$) and mean particles volume.

To replicate the growth of ice crystals, two opposite walls were moved at a given speed (Figure 1a), the speed of the ice front velocity measured experimentally. The simulation box was constraint to remain small in the two directions z and y, which allows us to neglect the curvature of ice crystals tips; they were therefore fixed at 50 $\mu$m. The y direction was thus extended to 100 $\mu$m to model enough particles to have statistical information. The other walls were maintained fix during the whole simulation, at a distance corresponding to particle radius, to allow the rotation of the particle placed in the box boundaries. Software limitations do not allow a periodic condition with rigid body assemblies. Considering the low surface charge of BN particles in water[19] and the large size of the particles, only the repulsive part of the Lennard-Jones potential was used to model particle interaction:

$$V(r) = \epsilon_0 (\frac{\sigma}{r})^{12}$$

where $\sigma$ = 1 $\mu$m is the particle diameter, $r$ the distance between two particles and $\varepsilon_0$ the Hamaker constant of the BN / water / BN interfaces. This potential, because it is only repulsive, creates an excluded volume to avoid unphysical particle overlaps, while simultaneously saving CPU time. The particle / ice front interaction is represented by a Van der Waals potential for a sphere and a plane near contact[20]:

$$W(h) = -\frac{AR}{6h}$$

where $R = \sigma / 2$ corresponds to the particle radius, $A$ the Hamaker constant of particle / water / ice interaction and $h$ the distance between the ice front and the particle. The non-retarded lifshitz theory was used to calculate the different Hamaker constants with the refractive index and dielectric constant of each material (hexagonal BN, water and ice). The input values for simulation were $H = 4 \times 10^{-20}$ J and $A = -3 \times 10^{-21}$ J. The negative value of the Hamaker constant for ice / particle interaction indicates repulsion, characteristic of many solidification front[21–23]. The simulations were carried out in an NVT ensemble with the temperature regulated by a Langevin thermostat. The Langevin thermostat allowed the addition of a viscous force characteristic of the solvent interaction, which is intimately linked to the repulsion / engulfment phenomenon.

The repulsive interaction between the ice front and the particle lead to a disjoining pressure thus to the appearance of a thin liquid film at the interface. If the particle comes closer to the front, there is a chance of being encapsulated in the ice. No existing model grasps the complexity of the repulsion / engulfment equilibrium in concentrated colloids and a simplified one introduced by Barr et al.[24] has been used here. It is based on the balance between drag and repulsive forces. The repulsive forces come from the interactions with the ice front and the drag forces from solvent interaction. The maximum drag force attained at the encapsulation can be calculated via Stokes's law:

$$F_{max}^d = 6\pi\eta R v_c \qquad (3)$$

where $\eta$ is the viscosity of water (1.8 mPa.s at 0°C), $R$ the particle radius and $v_c$ the critical velocity (determined experimentally at 35 $\mu$m/s for BN particle). The critical distance $\delta$ before encapsulation can be deduced by equating the force applied by the moving front (the opposite of the derivative of the interaction potential) with the maximum drag force undergoes by the particle: $-\frac{\partial W}{\partial h} = F_{max}^d$. The construction of the anisotropic particles introduced here let us apply all the interaction and forces on each individual sphere without any assumption on the whole particle. This model comprises up to 800 particles, thus close to 42000 individual spheres.

For visualization and image analysis purpose, the center, radius and normal vector of each particle was extracted and a plugin for Fiji was developed to visually reconstruct the particles from the knowledge of their geometry. The time-lapse density measurements were made by the use of a programmed plugin and the percolation region by the existing "Find connected region" plugin with Fiji. The images presented here have been made with the software Paraview.

3. Results and Discussion

3.1 Ice Growth Induced Self-Assembly of Anisotropic, Facetted Particles

The self-assembly of facetted particles during ice-templating occurs at length scale of tens of microns. The resulting spatial distribution is intimately linked to dynamical interactions between the particles and the growing crystals but also to viscous interactions with the unsolidified solvent. The imaging possibility offered now by synchrotron radiation facilities[25] allows the visualization of the growth of large crystals in real time[26] but the spatial resolution is still not sufficient to image *in situ* the dynamic arrangement of particles. After freezing and ice removal however, the resulting samples can be imaged at a resolution good enough to distinguish the particles. In the present paper, we have used high resolution KB focused X-Ray holo-tomography on impregnated samples of 1x1x1 mm$^3$ at the ID22NI beam line at the ESRF.

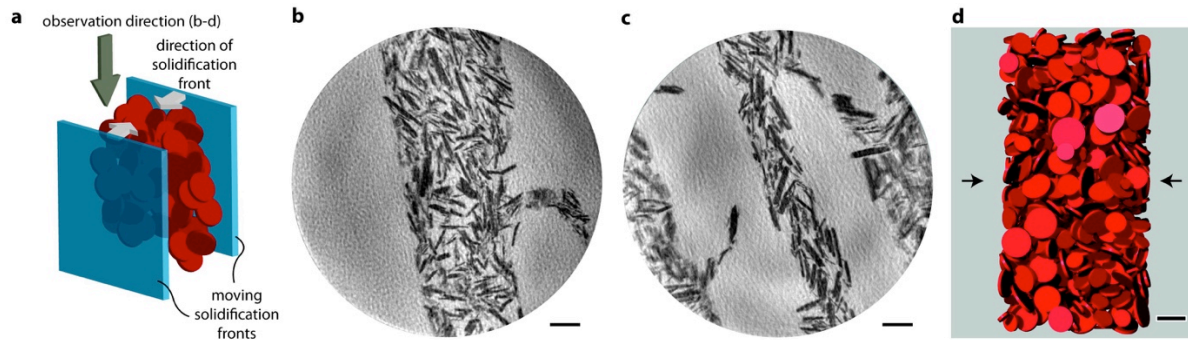

**Figure 1**. Comparison between the modeling initial state with 700 platelets and the holo-tomography reconstruction of ice templated samples, with the same viewing direction. (a) Schematic view of the sample geometry, the solidification fronts, their direction, and the observation direction for (b-d) (b) and (c) Cross-sections perpendicular to the wall direction (thus parallel to the lateral growth of the ice crystals) of holo-tomography reconstructions, as defined in (a). Structures obtained with the slow freezing rate (15 $\mu$m/s) (b) and the fast cooling rate (25 $\mu$m/s) (c), resulting respectively in the thickest and the thinnest wall. The two types of structure are visible, with a better alignment of particles in the outer regions than in the core for (b) and the homogenous orientation of the particles in (c). (d) Initial state of the modeled system, representing the random orientation and polydispersity of the particles size before compaction. The black arrows indicate the direction of the solidification fronts. Scale bars: 10 $\mu$m.

Microstructures of ice-templated BN platelets infiltrated by a silicon rubber are shown in Figure 1a and 1b. The 50 nm resolution of the holo-tomography is sufficient to discern the platelets packing in the wall. In Figure 1b, the slow solidification interface velocity (7 µm/s) leads to thick walls (~30 µm) and the particles that are directly in contact with the growing ice are well aligned, parallel to it. Oppositely, in the inner region of the wall the particles are less ordered. This type of structure will be referred to as the *sandwich structure*. When the solidification interface is moving faster, (close to 25 µm/s), the wall obtained is significantly thinner (~15 µm, Figure 1c) and the packing and alignment of particles is improved. This type of structure will be referred to as the *compact structure*.

The model used in this study consists in a Langevin simulation of colloidal particles with different interaction potential, one for the particles/ solidification front interaction and one for the particles/particles interaction (see details in section 2). The initial state of the modeled platelets is depicted in Figure 1d and shows the Gaussian size distribution of the particles, their random orientation, as well as the width of the model box (e.g. 50 µm). These dimensions are close to the ones of the walls obtained in the silicones/boron nitride composites. This model comprises up to 800 platelets in a volume of 100 x 50 x 50 µm$^3$. The simulated volume is willingly limited to those dimensions, to reflect closely the actual experimental conditions (Figure 1) and to neglect the curvature of the ice crystals in the freezing direction. Because crystal nucleation (and therefore branching) cannot be taken into account in this type of model, the lateral dimension has been extended just enough to have a representative number of particles in a reasonable simulation time. The benefits of self-assembly of large particles in bulk materials are demonstrated here with the thermal properties of the BN/silicon rubber composite shown in Figure 1. Controlling the orientation of BN particles to take advantage of their anisotropic thermal conductivity could be highly beneficial in thermal management applications, where the heat needs to be taken away from the components in the most efficient manner. BN is a prime choice for ceramic materials that are used in thermal management systems. The crystalline anisotropy of BN can be used to obtain platelet-like particles which exhibit orthotropic thermal properties[27] that are 20 times lower in the *c*-axis (30 W/m/K) direction than in the perpendicular plane (600 W/m/K).

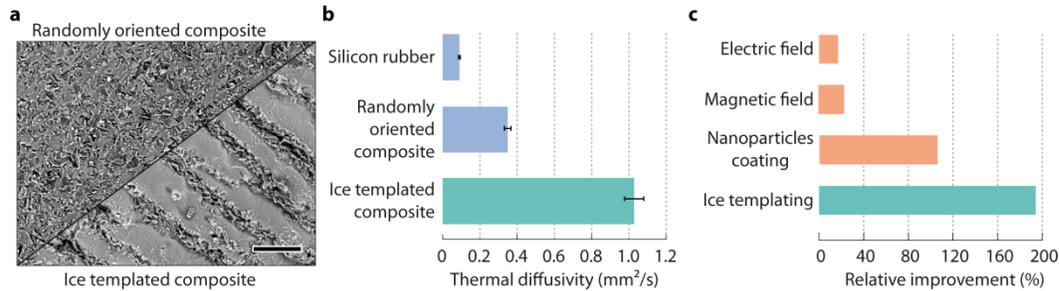

**Figure 2**. Improvement in thermal properties resulting from the control of the orientation of anisotropic particles. (a) Microstructure obtained for both randomly oriented (upper left) and ice templated Boron Nitride/silicon rubber composite (lower right). Scale bar: 100 $\mu$m. (b) Comparison of the thermal diffusivity of the two composites and the silicon rubber. (c) Comparison of the relative improvements RI (calculated as RI = ($\alpha_{aligned}$-$\alpha_{random}$)/ $\alpha_{random}$, where $\alpha_{aligned}$ and $\alpha_{random}$ are respectively the thermal diffusivities' of the materials with the aligned platelets and with the randomly oriented platelets) in thermal properties due to the alignment of BN particles induced by an electric field[28], a magnetic field (with[29] or without[30] superparamagnetic nanoparticles coating), or ice templating (this study). The BN volume fractions of these composites are respectively 0.6 vol.%[31], 10 vol.%[29], 5 vol.%[30], and 18 vol.%.

Here we took advantage of the ice growth induced self-assembly to obtain bulk, macroporous materials where the anisotropic particles are aligned along the freezing direction (Figure 2a). The samples were then infiltrated with a silicon rubber before their thermal properties were assessed. For the sake of comparison, a homogeneous composite was obtained by simply mixing the same amount of particles in the silicon rubber (Figure 2a). The corresponding microstructures are shown in Figure 2, along with the results of the thermal diffusivities measurements (Figure 2b). The material with the controlled architecture presents a thermal diffusivity almost three times greater than the homogeneous composite, illustrating the benefits of an aligned structure, even if the alignment is not completely perfect. Alternative strategies based on the application of an electric[28], magnetic (with the aid of superparamagnetic nanoparticles[29] or not[30]) field have been proposed to align BN particles to improve the thermal properties. The relative improvement in thermal properties (Figure 2c), estimated using the difference of the thermal properties in the random and aligned structures normalized by the thermal properties of the random structure, provides a comparison of the efficiency of these strategies. Ice templating is comparatively much more efficient. On top of that, it is also able to align particles in much more concentrated suspensions (almost 20 vol.%, compared to less than 5 vol.% for the other methods).

3.2 Predictions of Particle Packing

The driving force for ice growth induced self-assembly of large (>1 µm) particles is not random thermal motion. The potential energy needed to lift a particle (of 8 µm diameter and 0.5 µm thick) from its own diameter in water can be estimated by the following relationship:

$$U_p = m \cdot g \cdot \Delta h = \pi r^2 e (\rho_{BN} - \rho_{water}) \cdot g \cdot 2r = 2.6 \times 10^{-18} J \quad (1)$$

where $m$ is the weight of the particle, $r$ its diameter, $e$ its thickness, and $\rho_{BN}$ and $\rho_{water}$ the density of BN and water respectively. This energy is thus around three orders of magnitude superior to the energy brought by the thermal motion (of the order of $k_B T$, so $3.8 \times 10^{-21} J$ at 273K). This means that the behavior in water of particles of this size will not be controlled by thermal energy.

The common modeling approaches (e.g. Monte Carlo[32]) thus cannot accurately describe the behavior of the system. In particular, the particle encapsulation by the interface[22], a key feature of the phenomenon, cannot be taken into account. Discrete element modeling (DEM) is able to take into consideration several crucial aspects of the phenomena, such as the viscosity of the suspension, the interactions between particles (modeled here by a Van der Waals and Lennard-Jones pair potential for the ice and inter particles interactions respectively), and the continuous particle size distribution. The key feature implemented is the possible engulfment of particles by the moving solidification interface. Many different analytical models have been developed to predict this complex phenomenon[23,33–36], but are only valid for single, isolated particles. By using DEM, a simple force equilibrium between repulsion and drag forces can be introduced and thus take the dynamical and collective aspects (e.g. shock between particles) of this phenomena into account. It also provides access to the dynamics (time-lapse position and orientation) of particle redistribution by the growing ice crystals, which is currently impossible to obtain experimentally.

Particular attention must be paid to the particle shape and interactions. Different pair potentials can be used for anisotropic, facetted particles. These anisotropic potentials have been developed for clay suspension, such as the Gay-Berne[37,38] potential, based on Lennard-Jones potential and a quadripole orientation. The RE-squared[39] potential, based on an assembly of small isotropic Lennard-Jones potential has also been used. Even if those potentials allow a lot of the complexity involved in anisotropic particles interaction to be taken into account, they are limited to model systems with monomodal particles and therefore, cannot accurately describe the interactions with planar interface such as the solidification front here.

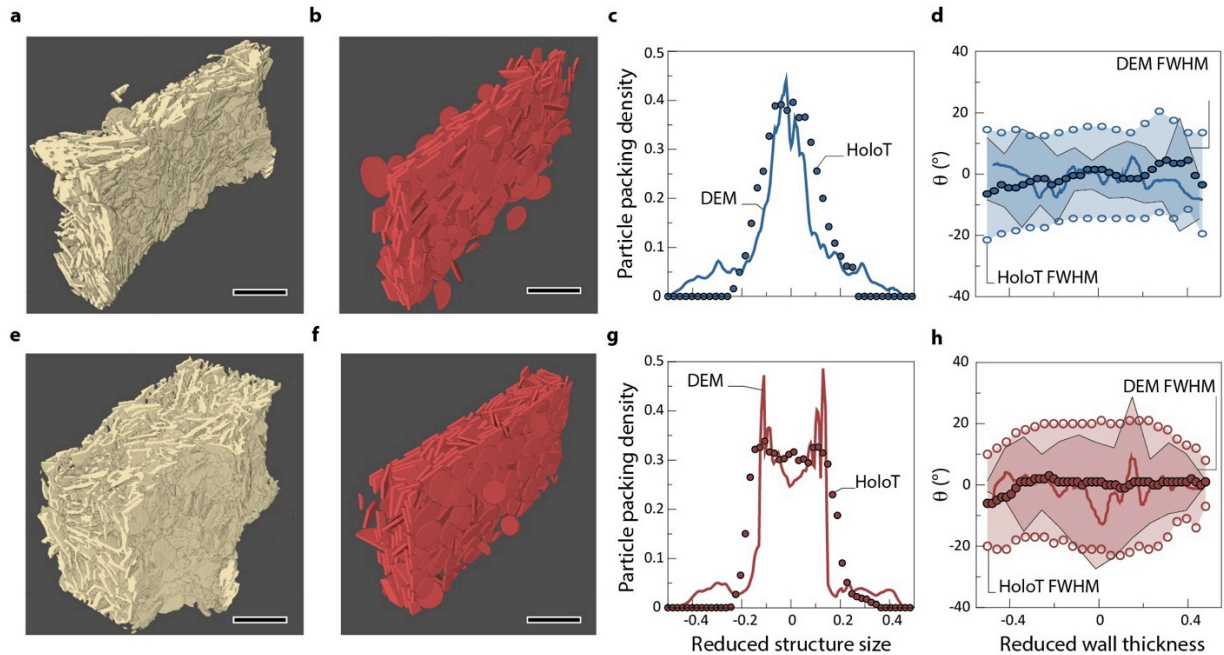

**Figure 3**. Comparison and characteristics of particle packing of the two types of architectures in the tomography reconstruction and the model. (a-d) Compact structure and (e-h) sandwich structure. (a) and (b), respectively (e) and (f) represent a 3D-view of the structures, as seen by holo-tomography (left) and DEM (right). (c) and (g) Packing density vs. reduced structure size. A continuous evolution of density is observed in the compact structure, while the density drops in the inner region in the sandwich structure. (d) and (h) Distribution of particles orientation for both structures vs. reduced wall thickness. The darker color indicates where the modeled and the experimental distributions overlap. Scale bar: 20 $\mu$m.

A different strategy has been developed here. Spheres (with an associated Lennard-Jones potential) are fixed together as cylindrical rigid bodies. The radius of those discs follows a Gaussian distribution to represent the polydispersity of platelet size. During the simulation, the software sums the interactions on all the spheres to calculate the resulting displacement and torque applied to the assembly. The simulated engulfment phenomenon can thus take place locally, at the scale of the constitutive spheres instead of the whole platelet and thus is closer to the reality. The comparison of the modeling and experimental results in terms of structures, particle density packing and distribution of orientations are shown in Figure 3. The predicted compact structures (Figure 3b) are comparable to the experimental one (Figure 3a), with highly oriented particles in a thin wall. The same conclusion can be drawn for the sandwich structures (Figure 3e and 3f), where the particles present a better alignment on the outer region of the wall than in the inner region. The use of reduced coordinates where the thickness of the wall is normalized by the periodicity $\lambda$ of the structure (structural wavelength[40]) allows the proper comparison between the experimental and predicted results.

The experimental and predicted particles packing (Figure 3c) are very similar: the density reaches a maximum value of 0.45 (0.41 for the experimental one) in the inner region of the wall. The high aspect ratio and the particle size distribution lower the packing density, compared to that of random close packing of monodispersed spheres (known to be about 0.64). For the sandwich structure, the evolution of the experimental and predicted density is again similar (Figure 3g). The density appears higher in the border and reaches a value of 0.45, similar to the compact structure while the density in the inner region falls to 0.25 (0.35 and 0.29 respectively for the actual structure). Both structures present approximately the same overall density (0.32).

The dispersion of particle orientations with respect to the wall direction is plotted in Figure 3d and Figure 3h, represented by their mean orientation and the full width at half maximum (FHWM) of their cumulative distribution. In the compact structure (Figure 3d), the mean orientation corresponds to that of the wall (mean angle of 0°), for both the model and the experiment. The concentration of the particle during the freezing does not induce any tilt of the particles. The FWHM is constant in the wall, and the value is slightly smaller in the model (10°) than in the experiment (15°). The particles are slightly more aligned in the model than in the experiment and therefore the density and alignment are also slightly overestimated. In the sandwich structure (Figure 3h), the FWHM varies from 10° in the outer region to 20° in the inner region, revealing the misalignment of the particles in the inner region.

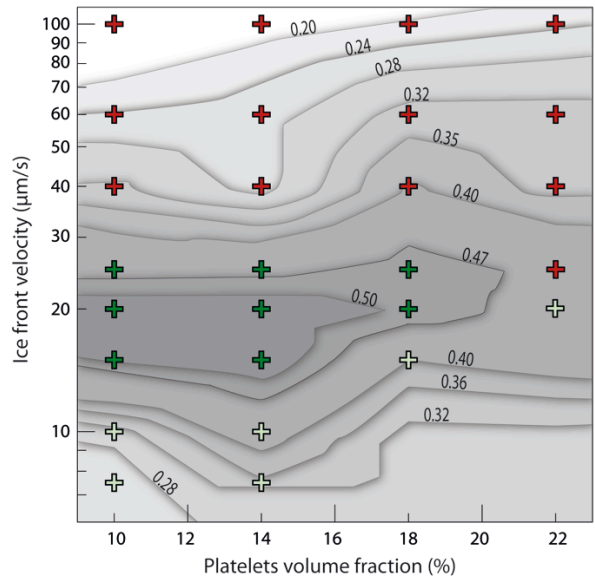

**Figure 4**. Phase diagram of the possible anisotropic particle assemblies. ✚ sandwich structure, ✚ compact structure, and ✚ engulfed structure (the encapsulation is defined as when more than 30% of the particles <u>initially present</u> have been engulfed in the ice). The background represents a map of the wall density predicted by DEM, each tone representing an isodensity region.

We can now analyze through a parametric study the resulting structure as a function of the experimental conditions. The predicted phase diagram is shown in Figure 4. This diagram is calculated using simulation, as a function of particle size and interface velocity, the two main experimental parameters that can be adjusted to control the resulting microstructure. Encapsulation is defined as the situation where 30% of the particles initially present have been encapsulated by the moving solidification front. The selection of this threshold of particle fraction as a criterion for encapsulation was made by looking at the fraction of engulfed particles at the critical velocities determined experimentally. To determine experimentally the critical velocity values, we froze the same suspension at different cooling rates and observed the resulting microstructure. The critical velocities values were determined when nearly all the particles did not rearrange with the front, in which case a lamellar structures pore/wall was not obtained anymore. Those structures are obtained at the same ice front velocity (above 40 µm/s) for the entire set of initial particle fractions (which confirmed this value as an intrinsic property of the particles) except for the highest solid loading (0.22). At this point the concentration is close to the percolation threshold (estimated at 0.215, see Experimental section and references[17,18] for more details). The transition between sandwich and compact structures occurs at a higher ice front velocity as the solid loading increases, as higher solid loadings impinge the movement of the particles. The percolating network of platelets restricts and ultimately suppresses particles rearrangement during the compaction by the solidification front.

The final densities are also represented in Figure 4 as an isodensity contour map. As described previously, the architecture of the wall reflects the particle orientation and thus the packing density. The packing density is calculated as the maximum particle density reached inside volume contained by the advancing solidification fronts. Area of high density corresponds to the compact structure. This phase diagram can be used as a predictive tool to adjust the properties of the final material, by predicting the density and structure of the particle packing as a function of the experimental conditions.

3.3 Dynamics: Self-Assembly Principles

Because the structures obtained after freezing are the result of the redistribution and concentration of particles by a solidification front, the observed differences arise from different interactions that take place in this peculiar system. In a first approximation, the process can be considered as equivalent to a simple powder compaction by two moving fronts. Nevertheless, particles can be encapsulated by the moving front, depending of the conditions (viscosity, velocity, existence of premelted films[23,33]). The dynamic behavior of the system can be investigated by DEM (Figure 5).

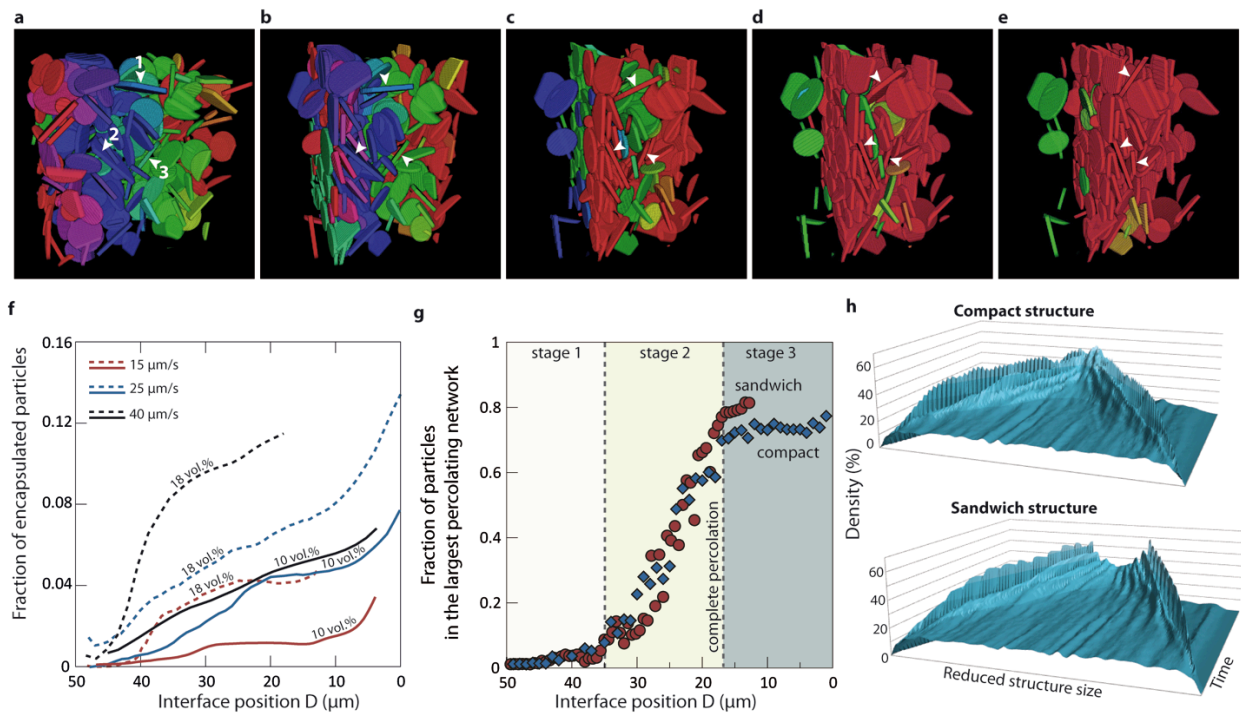

**Figure 5**. Dynamic evolution of particle organization and packing density. (a-e) Snapshots of the time-lapse evolution that illustrate the progressive alignment of the particles as the interface progresses. The color code corresponds to the percolation of particles: particles having the same colors belong to the same percolating network. (f) Dynamical evolution of the fraction of encapsulated particles during the freezing for two different volume fraction: 10 and 18 vol.% and three different interface velocities. (g) Evolution of the fraction of particles that belongs to the largest percolating network as a function of the interface position Δ, for the compact and sandwich structures. A complete percolation is obtained around 17 $\mu$m for both types of structure, although further local alignment of the particles occurs, as can be observed in (a-e). The fraction is not reaching a value of 1 as some particles are engulfed in the early stages of interface movement. (h) Time-lapse evolution of the density profile through the wall thickness, for both types of structure.

The time-lapse snapshots of the DEM results (Figure 5a to 5e) and the corresponding density profiles (Figure 5h) illustrate how the alignment is initiated in the outer region and then progressively propagates towards the inner region. Particles with the same color belong to the same percolating network. The criterion to identify the different regions is a simple pixel-growing algorithm, available in the image analysis software used (see Methods). A different RGB color is then attributed to each percolating network of particles. The evolution of the system follows three stages (Figure 5g). Because the interface moves slowly, very few or no particle are engulfed during the first moments of growth (stage 1), they are just locally aligned and densely packed (Figure 5a and 5b). Some are then engulfed (Figure 5c) as the interface keeps moving. The layer of aligned particles is still pushed and the concentration of particles in the inner region of the system starts to increase (Figure 5h). Percolating networks grow rapidly (stage 2). It starts to act against the displacement and rotation of particles. The pressure exerted by the moving interface is eventually not strong enough to overcome the counter pressure exerted by the percolating network of particles (stage 3); the system is ultimately jammed in a configuration where particles are aligned in the outer region and in the inner region (compact structure) or not (sandwich structure).

The evolution of the fraction of particles encapsulated in the ice is plotted in Figure 5f, for two initial concentrations (10 vol.% and 18 vol.%) and three different ice front velocities (15, 25 and 40 µm/s). For the 18 vol.% system, the density evolutions present similar trajectories at three different velocities (15, 25 and 40 µm/s), even if the final densities differ. The evolution is similar in a more dilute system (10 vol.%). An increase of either the volume fraction of particles or the interface velocity results in more encapsulation of particles by the moving interface.

Percolation arguments are not enough to predict the type of structure obtained. In both types of structure, complete percolation is obtained at approximately the same position of the moving interfaces (Figure 5g). Further local ordering of the particles is still possible after reaching complete percolation, as illustrated in Figure 5a to 5e (white arrows).

For large particles like the ones used here, the critical factor controlling the evolution of the system and the type of packing obtained is thus the encapsulation of the particles by the moving interface. With the same initial volume fraction of particles, an increase in interface velocity results in an increase of particle encapsulation. As a result, the concentration of the solid phase between the moving interfaces is lower. Particles have more freedom to reorient before the interface eventually encapsulates everything, leading to the compact structure. If all of them are repelled by the moving interface, the concentration is too high when the particle networks start to percolate. Particles do not have enough time to rearrange before total encapsulation, leading to the sandwich structure. The thinner the wall, the higher would be the shear stresses applied in the suspension for the same lateral growth velocity. With these conditions of solidification front velocity, particle size, and solid loading, the thickness of the wall nevertheless does not play an important role, since the shear forces in the vicinity of the solidification front induce ordering.

3.4 Self-Assembly of Cubic Facetted Particles

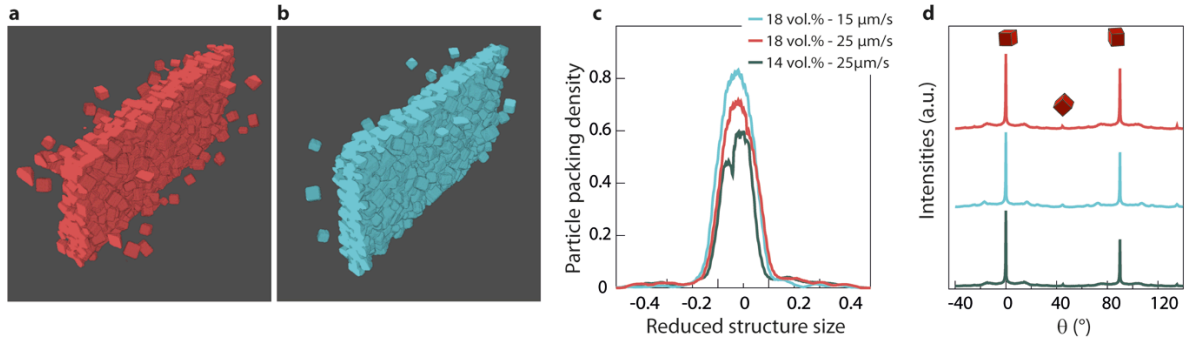

**Figure 6**. Ice growth induced self-assembly of cubic particles. (a) and (b) Final architectures predicted by DEM for 25 $\mu$m/s (a) and 15 $\mu$m/s (b) ice front velocity, at the same solid loading (18 vol.%). Denser packing's are obtained for the slowest front velocity. (c) Particle packing density for different concentrations (14 and 18 vol.%) and interface velocities (15 and 25 $\mu$m/s). (d) Orientation of the resulting architecture (see text for details) showing the alignment of the cubes along the wall direction (0° and 90°).

Being able to orient platelet-like particles in porous or bulk materials is highly valuable, in particular for structural materials but also piezoelectric or thermal management materials. However the self-assembly induced by ice-templating is not limited to this specific particle shape and might be extended to any type of facetted particle. To illustrate the fact that facetted particles can self-assemble even if they have an isotropic aspect ratio, the self-assembly of cubic particles has been explored with our simulation tool. Three different conditions (Figure 6) have been simulated, to predict the effect of the ice front velocity and the initial volume fraction. The obtained morphologies at two different ice front velocities are presented in Figure 6a and 6b. Only one type of organization is obtained with all the simulation conditions: after solidification, all the cubes are well aligned in the wall's direction. This is confirmed by the density profile in Figure 6c, where the maximum density is at the center of the wall. However, this value changes with the applied conditions. The higher the initial density and the slower the compaction rate, the higher the final density. The packing density can reach a value as high as 0.80, slightly higher than the theoretical random packing of cubes found in the literature (0.74 for reference[41], 0.78 for reference[42], both using different model type). The nearly perfect stacking in one direction (the direction of the wall thickness) can explain this slight discrepancy. The alignment of cubes is similar in all the conditions tested and the image analysis revealed nearly only the edges of the cube at 0° and 90° (Figure 6d). The cubes are then nearly all perfectly stacked in the thickness of the wall. Those three conditions reveal the potential of this model to easily determine the best conditions to obtain dense packing of cubic particles or any others shaped particles of interest.

4. Conclusions and Perspectives

Using the growth of ice crystals as a driving force for the self-assembly of large, facetted particles is an appealing alternative to the current evaporation or sedimentation strategies. A key conclusion of this work is that a high aspect ratio is not required for ice growth induced self-assembly, in contrast to the sequential deposition methods[7,9]. The steric interaction between the moving interface and the particle is highly localized; working with facetted particles is therefore the only morphological requirement to obtain alignment. The methodology developed here can easily be generalized to other systems, in particular by taking into account different models of interaction potentials for smaller particles where gravity is not the controlling parameter.

Although the particle alignment is not perfect, in comparison to superlattices obtained by sedimentation for instance, the benefits on the functional properties have been demonstrated here for thermal properties. The possibility to align particles in the bulk (centimeter sized samples) and in short times (minutes), holding the promise of novel crystalline materials with appealing functional properties, make it worth further research efforts.

AUTHOR INFORMATION

**Corresponding Author**

*E-mail: sylvain.deville@saint-gobain.com

**Funding Sources**

We acknowledge the financial support of the ANRT (Association Nationale Recherche Technologie) and Saint-Gobain through a CIFRE fellowship, convention #808/2010.

ACKNOWLEDGMENTS

We thank Loic Courtois and Michel Perez for introducing us to the LAMMPS package. We thank Peter Cloetens from the European Synchrotron Radiation Facility for helping us to carry out the holo-tomography experiment.